\documentclass[preprint]{elsarticle}
\usepackage{graphicx}   
\usepackage{amsmath,amssymb} 
\usepackage{hyperref}   

\title{Scalable Machine Learning Models for Predicting Quantum Transport in Disordered 2D Hexagonal Materials}
\author[iust]{Seyed Mahdi Mastoor \fnref{fn1} }

\author[iust]{Amirhossein Ahmadkhan Kordbacheh\corref{cor1}}
\cortext[cor1]{Corresponding author: \texttt{akordbacheh@iust.ac.ir}}

\address[iust]{Department of Physics, Iran University of Science and Technology, Tehran, Iran}
\fntext[fn1]{Email: \texttt{smahdimastoor@gmail.com}}

\date{}
\DeclareUnicodeCharacter{2212}{-}
\begin{document}
\begin{frontmatter}

\begin{abstract}
We introduce scalable machine learning models to predict two key electronic properties of disordered two-dimensional hexagonal nanomaterials: the transmission coefficient T(E) and the average local density of states (Average-LDOS). Using a tight-binding Hamiltonian combined with the Non-Equilibrium Green’s Function formalism, we generated a dataset of more than 400,000 unique nanoribbon configurations across graphene, germanene, silicene, and stanene with varying geometries, impurity concentrations, and energy levels. A central contribution of this work is the development of a geometry-driven and physically interpretable feature space that enables generalization across material classes and system sizes. We systematically evaluate Random Forest regression and classification models in terms of accuracy, stability, and extrapolation ability. Regression consistently outperforms classification in capturing continuous transport behavior on in-domain data, while extrapolation performance degrades, revealing the limitations of tree-based models in unseen regimes. This study demonstrates a data-driven and transferable framework for accelerating quantum transport prediction in 2D nanostructures with disorder, providing new insights into structure–property relationships and guiding future development of physics-informed learning models for materials science.

\end{abstract}

\begin{keyword}
Quantum transport \sep NEGF \sep Machine learning \sep 2D materials  
\end{keyword}

\end{frontmatter}

\newpage
\section{Introduction}
In recent years, research on two-dimensional (2D) materials has grown significantly due to their unique physical and electronic properties. These materials offer promising opportunities for the design and development of next-generation electronic and spintronic devices \cite{lin2023recent}. A wide variety of 2D materials can be created by stacking different monolayers to form heterostructures. Theoretical predictions suggest that nearly 5800 such materials exist, with around 1800 successfully synthesized in laboratories  \cite{Ares2022-wv} \cite{Mounet2018-wm}. Most 2D materials naturally adopt a hexagonal lattice structure, favored for its high stability and dense atomic packing \cite{Ding2019-ag}.The atomic scale thickness and remarkable transport properties of materials like graphene and germanene have made them central to advancements in nanoelectronics and spintronics\cite{Ahn2020-vp}\cite{Sharma2017-vm}.

There are different types of methods for investigating transport properties of materials. Among them, density functional theory (DFT) has been used widely for investigate electronic and spin properties. DFT cannot directly calculate transport properties; one of the methods for this aim is using the scattering matrix  approach (S-matrix)\cite{lesovik2011scattering}. But the S-matrix has constraint about non-equilibrium conditions. Non-equilibrium Green functions (NEGF) is an other method that can be used to study charge and spin transport. NEGF can handle out of equilibrium condition and also inelastic scattering\cite{Rungger2020-bq}\cite{Datta2012-sy}.
However, substantial computational resources are often required when these traditional methods are applied to large or disordered systems, as they become computationally expensive.

To address these limitations, recent research has introduced novel machine learning (ML) and deep learning (DL) techniques to predict transport properties more efficiently.
ML is becoming a powerful tool in the study of electron transport properties in nanoscale structures and materials with impurities and defects.
One of its notable applications is the prediction of conductivity and transport probabilities in low-dimensional systems without the need to solve complex quantum transport equations \cite{Fang2022-fv}
\cite{Jing2023-oy} \cite{Lopez-Bezanilla2014-xi}\cite{Ko2021-mr}. Developing a reliable and scalable model to predict physical quantities, such as transmission coefficients can be highly beneficial to researchers studying nanostructures and designing devices based on charge and spin transport \cite{Jia2021-ty}.

The random forest algorithm is reported as a suitable model to predict transmission coefficients in the case that a regression problem is converted to classification \cite{Ghosh2023-pc}. 
Since various 2D materials appear in hexagonal lattice shape \cite{Ding2019-ag}; In this work,  models are presented to predict the transmission and average local density of states (average-LDOS) of the hexagonal lattice. Models are made scalable by selecting a feature space related to lattice geometry that has not been considered before. Besides, we compare the efficiency of regression versus classification methods in predicting transmission and present the advantage and trade-off of each approach. The comparison seeks to guide researchers on choosing the right machine learning techniques for future quantum transport studies in low-dimensional systems.

\section{Methodology}

\subsection{Tight binding model and non-equilibrium Green's
function approach}
\begin{figure}
    \centering
    \includegraphics[width=1\linewidth]{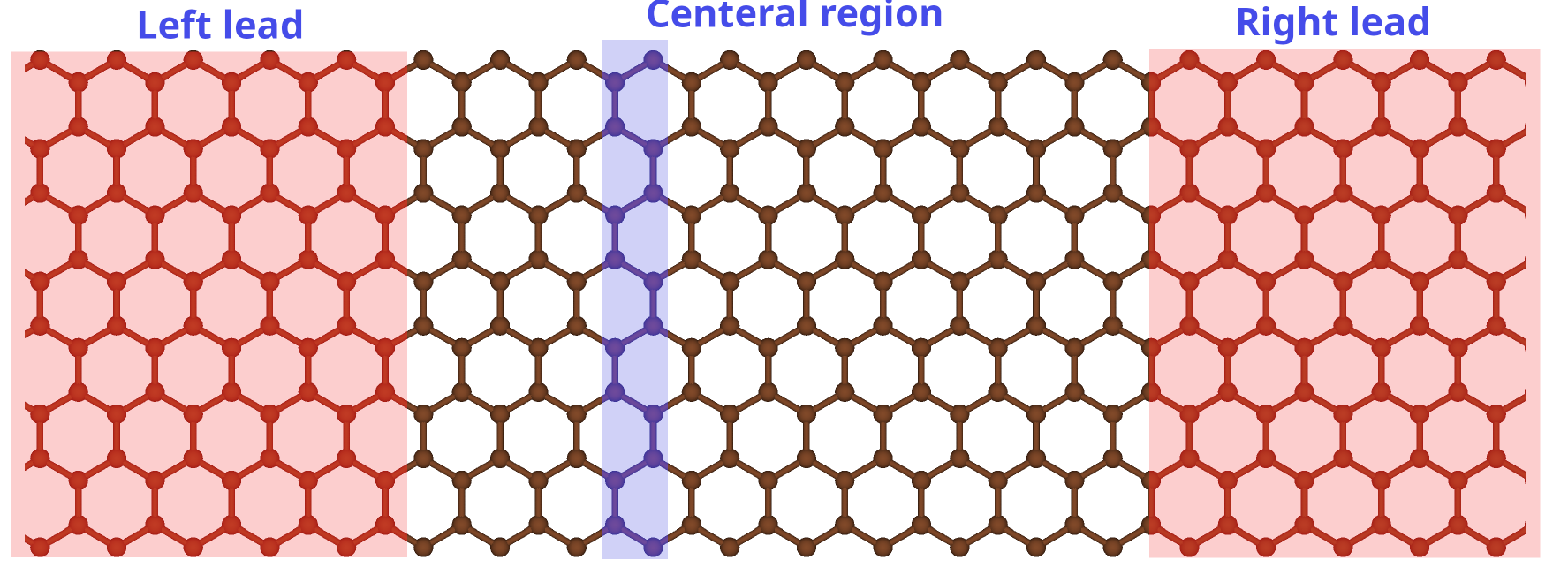}
    \caption{Schematic of the two terminal device. Red color region represent leads and The unit cell is highlighted in blue.  }
    \label{fig:shematic}
\end{figure}

We consider two-terminal device configurations, where the central scattering region contains magnetic impurities and is connected to nonmagnetic leads on each sides (fig.\ref{fig:shematic}). The system is modeled using a tight-binding Hamiltonian of the form:
\begin{equation}
    H=\sum_i \varepsilon_{i\sigma}\space C_{i\sigma}C_{i\sigma}^\dagger 
    \space 
    + 
    \sum_{<ij>\sigma\sigma^\prime}
    t_{ij}^{\sigma\sigma^\prime}
    C_{j\sigma^\prime}C_{i\sigma}^\dagger
\end{equation}
where $\varepsilon$  is the on-site energy and equal to $m_i\Delta\sigma_z$ , where $m_i \in [-1,+1] $ denotes the local magnetic direction randomly assigned to the site $i$. Also, $\Delta$ as the exchange energy set to unit. $t_{ij}^{\sigma\sigma^\prime}$
present the hopping amplitude between sites 
$i$ and 
$j$.

Graphene, germanene, silicene, and stanene, four of the most prominent 2D hexagonal lattice materials are used to construct the dataset. These materials are chosen for their well-documented electronic properties, structural similarity and relevance to emerging nanostructured device applications \cite{Le2014-ia}\cite{zhao2014spin}\cite{Lyu2019-tb}. Including a diverse set of materials ensures that the trained models can generalize across a broader class of 2D hexagonal systems. For each material, numerous device configurations are calculated under varying physical conditions, such as geometry size and impurity concentration. This allows us to construct a comprehensive dataset of average-LDOS and transmission coefficients, in a position to capture the highly complex transport behavior inherent in hexagonal lattice structures.
The hopping parameters are set based on material-specific values: $-2.6$ eV for graphene, $-1.3$ eV for germanene, $-1.6$ eV for silicene, and $-0.79$  eV for stanene. \cite{Ezawa2015-bf}\cite{Li2008-zy}\cite{Rahmani_Ivriq2019-uf}\cite{Liu2011-ae}\cite{Faisal2025-ku}. The model operates through a straightforward tight-binding method which examines how geometry and disorder influence transport properties while spin–orbit coupling and buckling, both of which can influence the transport properties, omitted.

In order to establish that our models can generalize across many realistic device configurations, the geometry of the scattering region was varied in two ways: the number of atoms per unit cell was changed from 6 to 32, and the number of unit cells in the transport direction was changed from 1 to 7. The ranges we use yield narrow and wide nanoribbons as well as short and long devices within the realistic size ranges used as designs in nanoelectronics. We also randomly distributed magnetic impurities within the scattering region at a concentration of 0\% to 10\% of the total number of atomic sites. These ranges of concentration range from pristine systems to those with moderate disorder, which allows the models to learn how the impurity-induced scattering affects transport processes. Altogether, the selections we have made provide a physically meaningful dataset that covers a wide range and enables the models to learn from an adequate set of training data that will be useful for scalable machine learning.

For each configuration, we compute two physical quantities that characterize the transport behavior: the transmission coefficient
$T(E)$ and Average-LDOS , both calculated using the NEGF formalism. $T(E)$ is calculated as:
\begin{equation}
    T(E)=Trace[\Gamma_1G^R\Gamma_2G^A]
\end{equation}
and Average-LDOS : 
\begin{equation}
    Average -LDOS=\frac{1}{N}\sum_1 ^N LDOS_i 
\end{equation}
where $LDOS_i=\frac{1}{\pi}Im[G_i(E)]$ is local density of state per site and $G^{R,A}=[E-H-\Sigma^{R,A}_1-\Sigma^{R,A}_2]^{-1}$ is the retarded/advanced Green's functions of the scattering
region and $\Sigma^{R,A}_{1,2}$ being the retarded/advanced self energy of the left/
right electrode, which can be numerically evaluated using an iterative recursive algorithm \cite{Sancho1984-dp}, and it's worth to note that $\Sigma^{A}_{1,2} =(\Sigma^{R}_{1,2})^\dagger $.  broadening matrix due to strong couplings
between the central region and left (right) lead of the
system is defined as $\Gamma_{1,2}=i[\Sigma^{R}_{1,2}-\Sigma^{A}_{1,2}]$.Also  the number of open channels in leads expressed as $M(E)_{1,2} = \dfrac{1}{2\pi}Tr[\Gamma_{1,2}]$ , because 
$\Gamma$ has a nonzero trace only for propagating modes.

\subsection{Machine learning}
Physically meaningful parameters that specify the geometry and transport properties of the system are used to build the feature space. These collectively define the size and shape of the system and include the scattering region's width and length , where $width = \dfrac{w}{2} \times \dfrac{\sqrt{3}}{2}a_0$ and $length = (l-1)a_0 +\dfrac{a_0}{2}$; $w$ is number of atoms per unit cell and $l$ is number of unit cells, and $a_0$ is bond length corresponds to the lattice constant and set to : $1.42 \text{\AA}$ for graphene \cite{Yang2018-ob}, $2.83 \text{\AA}$ for stanene \cite{Khan2017-lg}, $2.46 \text{\AA}$ for germanene \cite{Chen2016-gj} and $2.35 \text{\AA}$ for silicene \cite{Yamada-Takamura2014-tr}; for simplicity they divided to $\text{\AA}$ in dataset. Total number of atoms ($N$) as it used in Average-LDOS,  and the number of magnetic impurities ($nm$) are also included. We also take into account the hopping parameter, which varies depending on the material and the target energy level at which the transmission $T(E)$ and Average-LDOS are calculated normalized by dived to the hopping parameter, $E=E/|t|$.This normalization prevents material identity from being entangled with the energy scale and enables meaningful comparison across materials with different bandwidths.Also material-specific differences are encoded not only via the hopping parameter and normalized energy but also through the geometry derived from each material’s lattice constant. This feature set enables the models to be general and scalable across various materials and configurations while capturing the fundamental physics governing quantum transport. 

Random Forest (RF) classification and regression algorithms are employed in this research. RF is an ensemble learning method where numerous decision trees are generated during training and outputs are obtained from them, and is extremely powerful and reliable for most prediction tasks \cite{Breiman2001-tm}. Among the greatest strengths of Random Forest models is that it has the ability to learn high-order nonlinear dependencies between features and target variables\cite{Auret2012-st}.

RF models are extremely scalable and effective when dealing with large datasets. Their parallelizability and the fact that they are inherently resilient to overfitting also mean that they are very effective at dealing with high-dimensional input spaces as well as noisy data \cite{Genuer2017-xc}. In the application of quantum transport data, for instance, this characteristic makes RF highly suitable where fine features are governed by very many interactive parameters such as geometry, disorder, and energy levels.
Apart from Random Forest (RF), we also used Multilayer Perceptron (MLP) and Support Vector (SV) models to check their learning capacity of the complex and nonlinear relationships behind quantum transport data. MLP was chosen due to its capacity to map complex non-linear functions onto layered neural structure \cite{Amor2021-in}, and SV was included as a kernel-based model that is well known for its proficiency in handling high-dimensional and continuous feature spaces\cite{D2024-cd}. Both models have been successfully applied in physical and material modeling experiments\cite{Owolabi2021-zp}\cite{Rajan2018-qs}\cite{Zhang2021-gq}. The MLP and SVR models delivered acceptable results in our study but their performance numbers did not reach the level of the Random Forest (RF) model. Table \ref{select model} summarizes the cross-validated metrics for all three models, obtained using a subset of the dataset and evaluated through GroupKFold cross-validation. The results from the pre-testing phase demonstrate that the RF model delivers the most accurate and efficient mapping between structural and transport parameters in our dataset. The Random Forest (RF) model served as the base model for all future activities. These activities included evaluation, visualization, and interpretation.

\begin{table}
    \centering
    \resizebox{\textwidth}{!}{
    \begin{tabular}{|c|c|c|c|c|c|c|}\hline
         Model&  MAE-mean&  RMSE-mean&  R2-mean&  Accuracy-mean&  F1-mean& Time-mean-s\\\hline
         RF&  0.215&  0.414&  0.994&  0.821&  0.818& 3.931\\\hline
         MLP&  0.691&  0.944&  0.971&  0.491&  0.534& 115.057\\\hline
         SVR&  3.633&  4.846&  4.841&  0.125&  0.168& 153.669\\ \hline
    \end{tabular}
    }
    \caption{Cross-validated performance comparison of Random Forest (RF), Multi-Layer Perceptron (MLP), and Support Vector Regressor (SVR) models on a representative subset of the dataset using GroupKFold (k=5). Reported values represent the mean performance across folds, demonstrating that the RF model consistently outperforms the others and is thus selected as the primary model for further analysis.}
    \label{select model}
\end{table}

Performance of our machine learning models is assessed using popularly employed evaluation metrics. For regression tasks, we employ the Mean Absolute Error (MAE), the Root Mean Square Error (RMSE), and the coefficient of determination (R²). These metrics help quantify how closely the predicted values match the actual continuous outputs, with lower MAE and RMSE indicating better performance, and R² indicating how well the model explains the variance in the data.

For classification tasks, we evaluate model performance using accuracy and the F1-score. Accuracy measures the proportion of correctly predicted labels, while the F1-score which is the harmonic mean of precision and recall, provides a balanced view of the model’s ability to classify each class correctly, especially useful when class distributions are imbalanced.

To prevent overfitting and ensure true generalization across different device configurations, group-wise k-fold cross-validation was employed rather than standard random splitting. In dataset, each device (defined by its geometric parameters and impurity configuration) contains multiple energy points. Using random k-fold split could lead to samples from the same physical device appearing in both the training and test sets, causing data leakage and artificially high performance. By applying GroupKFold, where grouping is based on device identity, we make sure that all energy points coming from a given device are strictly in either the training or validation set to give a more reliable and unbiased estimate of model performance \cite{Wang2022-hv}. Additionally, GridSearch is also used to systematically explore different combinations of hyperparameters (such as the number of estimators, tree depth, and minimum samples per split) and then select the optimal configuration \cite{Probst2019-xd}.

Furthermore, polynomial feature expansion is incorporated to capture nonlinear relationships between input features and target variables \cite{Maertens2006-na}. By transforming the original feature space onto interaction terms and higher order components, we enable the models particularly Random Forest to learn the complex dependencies in our dataset better, which is essential given the nonlinearity of quantum transport phenomena in disordered systems.

\section{Results}
The dataset has been created by calculating T(E) and Average-LDOS for various shapes of scatter regions, as mentioned by the NEGF method. All nanoribbons were built with zigzag edges because these edges generate unique edge-localized states, and transport channels, which differ from armchair configurations \cite{Peres2006-bp} \cite{Klaassen2025-db}. The transport quantization and magnetic properties of 2D hexagonal materials strongly depend on the zigzag termination patterns. The energy range chosen for each material is by considering the transport energy window on the basis of the band structure of the that material. Energy range was set to approximately $[-2.5,2.5]$ eV for graphene \cite{Lherbier2012-nu}to span its Dirac cone linear dispersion region, and $[−1.5,1.5]$ eV for germanene \cite{Liu2022-pm}, stanene \cite{Lu2017-xw}, and silicene \cite{Voon2015-wl}, encompassing their symmetric conduction and valence bands near the Fermi level.
So, the $4.5 \times 10^5$ sample was calculated, each corresponding to a unique configuration defined by the size of the system, the concentration of impurities, the type of material and the energy. 
The dataset is divided  into train and test, where the test data are $20 \%  $ of the whole. 

\begin{figure}[]
     \centering
     \includegraphics[width=1 \linewidth]{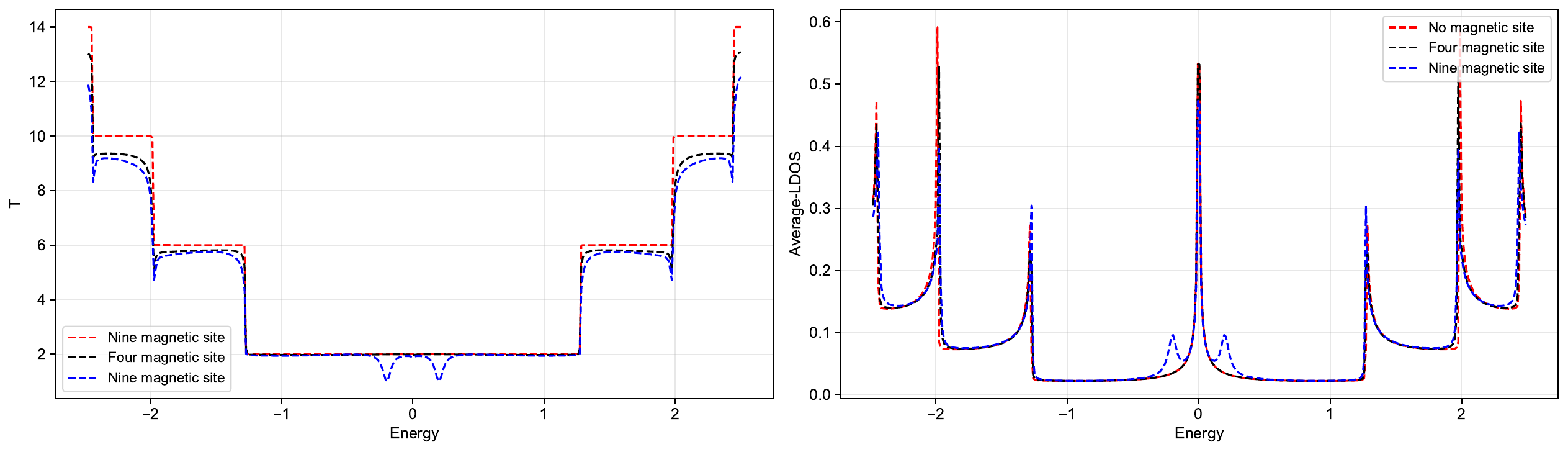}
     \caption{The transmission (T) and average local density of states (LDOS) are plotted as a function of energy (eV) for three random magnetic configurations of a graphene nanoribbon (width: 9.83 nm, length: 9.23 nm).}
     \label{fig:1}
 \end{figure}

\begin{figure}
     \centering
     \includegraphics[width=1\linewidth]{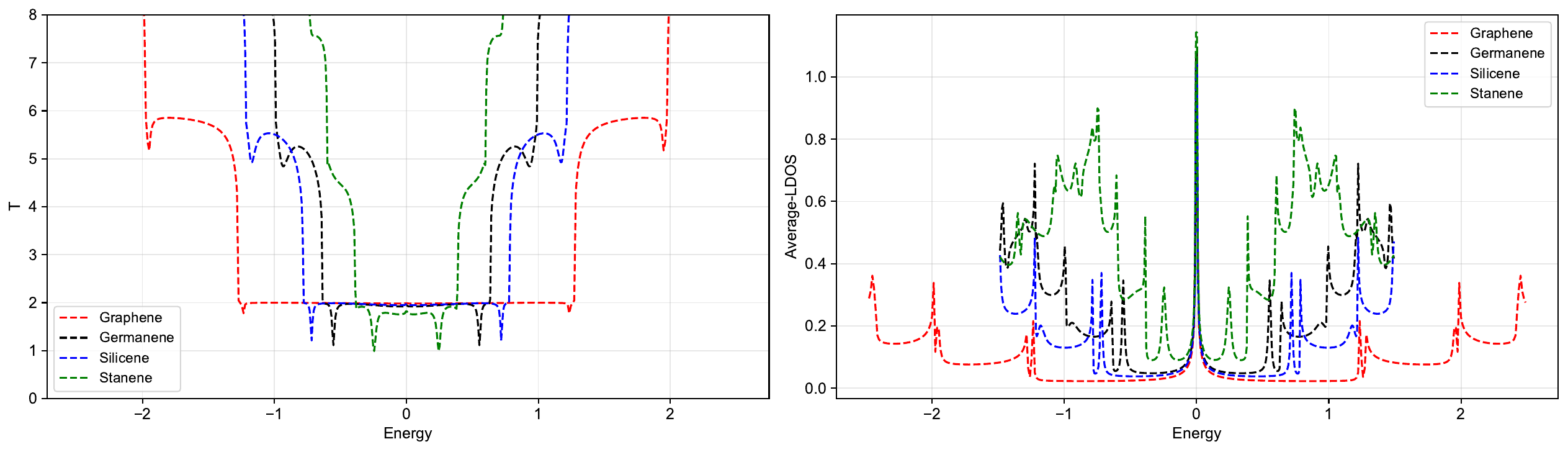}
     \caption{Comparison of 
T(E) and Average-LDOS for materials with 
equal number of sites $N=48$ and $nm=4$ used in the training dataset. }
     \label{fig:4m}
 \end{figure}

Magnetic impurities, which were randomly distributed within the scattering region, introduce disorder into the system. The disorder generates complex quantum interference effects, making the transport quantities display very nonlinear behavior \cite{Visuri2023-mh}. Nonlinearities are contained in the dataset and pose a challenging prediction task, as illustrated in fig.\ref{fig:1}. Comparison of the same configuration with equal number of magnetic impurities for graphene, germanene, silicene, and stanene; (fig.\ref{fig:4m}) show that T(E) and Average-LDOS exhibit distinct behaviors. With an increase in the width of the system, there is a greater number of quantized steps in the transmission spectrum as a function of energy. This is the characteristic behavior of wider systems that support more conducting channels. On the other hand, the introduction of magnetic impurities tends to smooth out these transmission steps by introducing scattering centers which disrupt coherent transport \cite{Kordbacheh2014-bd}\cite{Rahmani_Ivriq2019-uf}. Furthermore, impurities reduce the overall magnitude of 
T(E) and the average-LDOS near the Fermi energy, leading to a suppression of transport characteristics. In addition, a sharp peak appears in the average-LDOS around 
$E=0$, which can be attributed to the presence of edge or localized states. These results highlight the sensitivity of quantum transport properties to both disorder and geometrical parameters of the system \cite{Shimomura2011-nt}\cite{Zerah-Harush2020-qz}.

By considering our desired features, we have 6 inputs.
Polynomial features used to capture nonlinear relationships between input features and target variables. Since transmission and average-LDOS in hexagonal lattices are affected by complex quantum interference and impurity scattering\cite{Sukhachov2020-cc}\cite{Kordbacheh2013-tf}, their dependence on system parameters is often nonlinear. The transport properties are not simply proportional to the lattice size or impurity concentration.

For each target, we train distinct models because of differences in values and their distributions. Since transmission and average-LDOS are continuous variables, a regression model is the natural choice. However,classification models require discrete labels. we discretized targets values by using $C=round[y]$ , where C present classes. Convert continuous values into categories, making classification possible.
 To have a fair comparison, grid space of values for searching the best hyperparameters has been set equal for all models and optimized parameters selected from them. To avoid overfitting, the GroupKFold, method by setting k to 5 has been applied. Table \ref{tab:hyperparameters} shows results of the finetuning of models.   
 \begin{table}[htbp]
    \centering
    \resizebox{\textwidth}{!}{ 
        \begin{tabular}{|c|c|c|c|c|} 
            \hline 
            \textbf{Model} & \textbf{n-estimators} & \textbf{max-depth} & \textbf{min-samples-split} & \textbf{PolynomialFeatures-degree} \\ 
            \hline 
            Model-1 & 200& 20& 2 & 3 \\ 
            \hline 
            Model-2 & 200 & 20& 2& 3 \\ 
            \hline 
            Model-3 & 200 & 20 & 2 & 3\\ 
            \hline 
            Model-4 & 200& 20 & 2 & 3 \\ 
            \hline
        \end{tabular}
    }
    \caption{
        Determined hyperparameters by grid search method for:
        Model-1: RF regression for $T$,
        Model-2: RF classifier for $T$,
        Model-3: RF regression for average-LDOS,
        Model-4: RF classifier for average-LDOS.
    }
    \label{tab:hyperparameters}
\end{table}

We used scikit-learn to train and evaluate performance \cite{scikit-learn}. 
\begin{figure}
    \centering
    \includegraphics[width=1\linewidth]{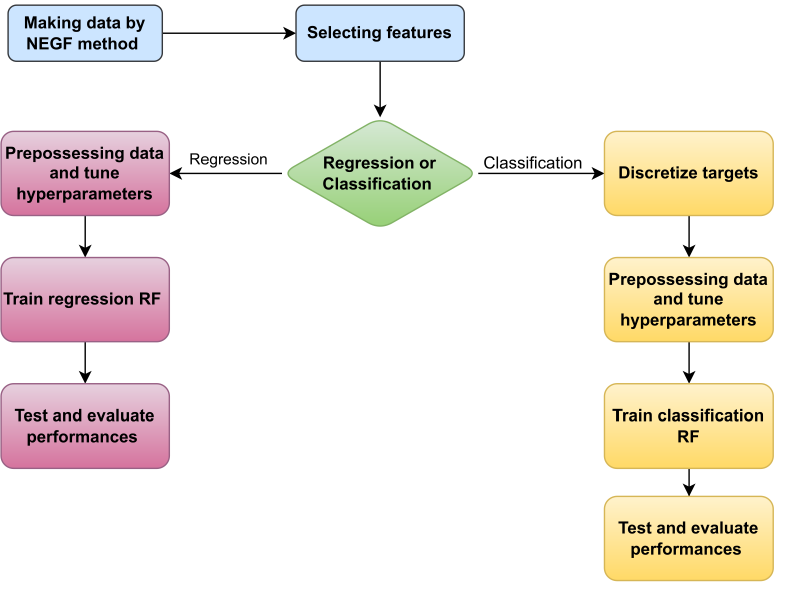}
    \caption{Flowchart of how we train and evaluate models }
    \label{fig:flowchart}
\end{figure}
fig.\ref{fig:flowchart} shows steps of our work as a flowchart.
The tuned hyperparameters are the number of decision trees (n-estimators), the maximum depth of each tree (max-depth), the minimum number of samples required to split a node (min-samples-split), and the degree of polynomial features to include for input expansion. By selecting these parameters properly, we ensure that each model is properly fitted to capture the complex patterns in the data without overfitting. Every model was enhanced by adding third-degree polynomial features.

By training models on training data, the performance of the models has been evaluated, also constraints $0 \le T(E) \le M(E)$ and average-LDOS $ \ge 0$ were imposed to ensure physical consistency and prevent unphysical results. Table \ref{tab:performance models of T} shows the performance to predict transmission values as the target and table \ref{tab:performance models of LDOS} for average-LDOS as the target.
\begin{table}
    \centering
     \resizebox{\textwidth}{!}{
    \begin{tabular}{|c|c|c|c|c|c|} \hline 
         \textbf{Model}&  \textbf{MAE}&  \textbf{RMSE}&  \textbf{$R^2$} &  \textbf{Accuracy}& \textbf{F1}\\ \hline 
         Model-1&  0.029&  0.091&  0.999&  0.971& 0.971\\\hline
   Model-1*& 0.042& 0.162& 0.999& 0.959&0.959\\\hline 
         Model-2&  0.032&  0.191&  0.9980&  0.968& 0.968\\ \hline
 Model-2*& 0.047& 0.256& 0.998& 0.959&0.959\\ \hline
    \end{tabular}
   }
    \caption{Performance of models in predicting 
$T$ on the test data. Models marked with an asterisk (*) were trained and tested without polynomial feature expansion.}
    \label{tab:performance models of T}
    
\end{table}
\begin{table}
    \centering
     \resizebox{\textwidth}{!}{
    \begin{tabular}{|c|c|c|c|c|c|} \hline 
         \textbf{Model}&  \textbf{MAE}&  \textbf{RMSE}&  \textbf{$R^2$} &  \textbf{Accuracy}& \textbf{F1}\\ \hline 
         Model-3&  0.006&  0.0434&  0.964&  0.993& 0.992\\\hline
 Model-3*& 0.013& 0.068& 0.911& 0.986&0.986\\\hline 
         Model-4&  0.064&  0.511&  0.954&  0.954& 0.953\\ \hline
 Model-4*& 0.141& 0.809& 0.879& 0.925&0.924\\ \hline
    \end{tabular}
   }
    \caption{Performance of models in predicting 
verage-LDOS on the test data. Models marked with an asterisk (*) were trained and tested without polynomial feature expansion.}
    \label{tab:performance models of LDOS}
    
\end{table}
MAE, RMSE and $R^2$ typically used to measure performance of regression models. As we compare classification versus regression we should use the same measure metric, So, for calculating accuracy as how many classes predictions were correct out of all predictions and F1 that computes the average of precision and recall, continues predicted values of regression were converted  by rounding to classes. On the other hand, to compute regression metrics for classification task, we consider classes that are produced by rounding as actual values of targets. 

Table \ref{tab:performance models of T} shows that the regression model significantly outperforms the classification model for predicting
$T(E)$. The regression model achieves an MAE
score of 0.029, compared to the classification model’s MAE of 0.032; indicating more precise predictions. Also, the lower accuracy and F1-score for classification indicate that rounding leads to a loss of precision in capturing variations in
$T$. The inherent capabilities of tree-based ensembles to capture non-linear and high-order feature interactions motivate an additional experiment without polynomial feature expansion in order to discern its actual impact on model performance. Models with polynomial expanded features showed slightly better performance compared to models without expansion, Table \ref{tab:performance models of T}. This could be explained by the fact that polynomial expansion reshapes the input feature distribution, thus allowing the ensemble to explore a richer partitioning of the feature space and capture smoother probabilistic relationships between geometric parameters and transport responses.

Similarly, Table \ref{tab:performance models of LDOS} shows that regression is also superior for average-LDOS predictions, with 
MAE = 0.006, whereas classification achieves an MAE of 0.064. The classification model struggles with capturing finer variations in average-LDOS due to discretizations effects. Discretizing continuous values by rounding leads to information loss and this approach affects the performance of RF classification.
Classification forces continuous values into discrete bins, making it unable to capture small variations in 
$T$ and average-LDOS that regression can learn directly. Additionally, for predicting the average-LDOS, models with polynomially expanded features also exhibited superior performance compared to those without feature expansion.
\begin{figure}
    \centering
    \includegraphics[width=1\linewidth]{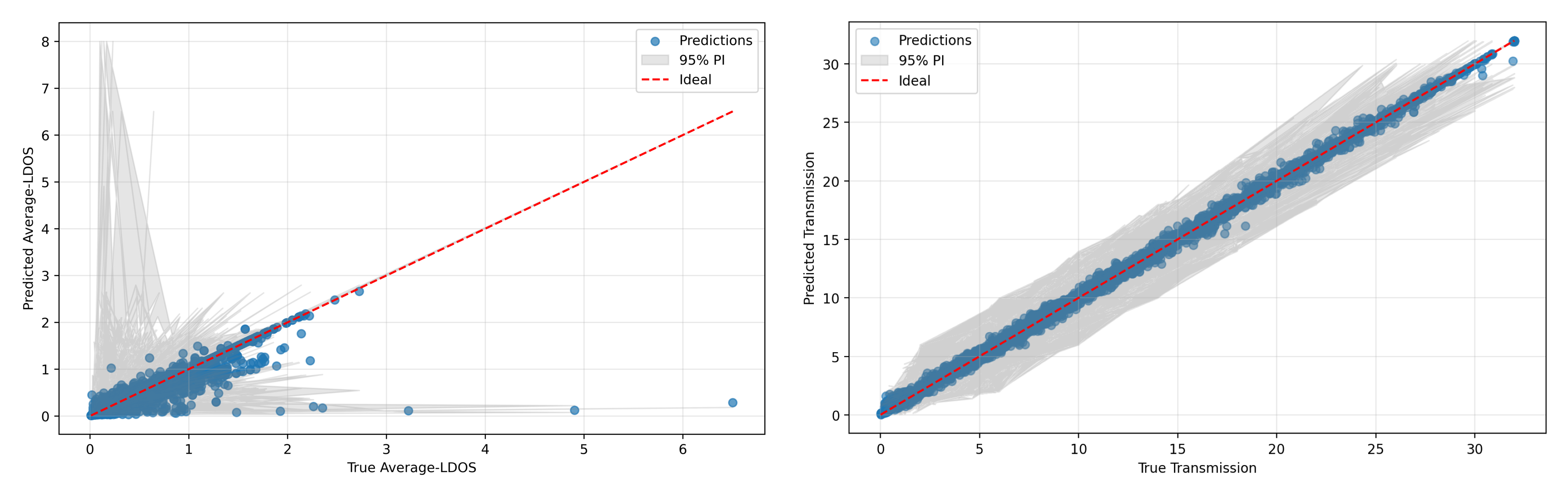}
    \caption{Regression model performance. Comparing the predicted vs. true values for Average-LDOS and Transmission ($T$). The dashed red line represents the ideal prediction ($y=x$), and the shaded area shows the $95\%$ Prediction Interval (PI).}
    \label{fig:pi}
\end{figure}

\begin{figure}
    \centering
    \includegraphics[width=1\linewidth]{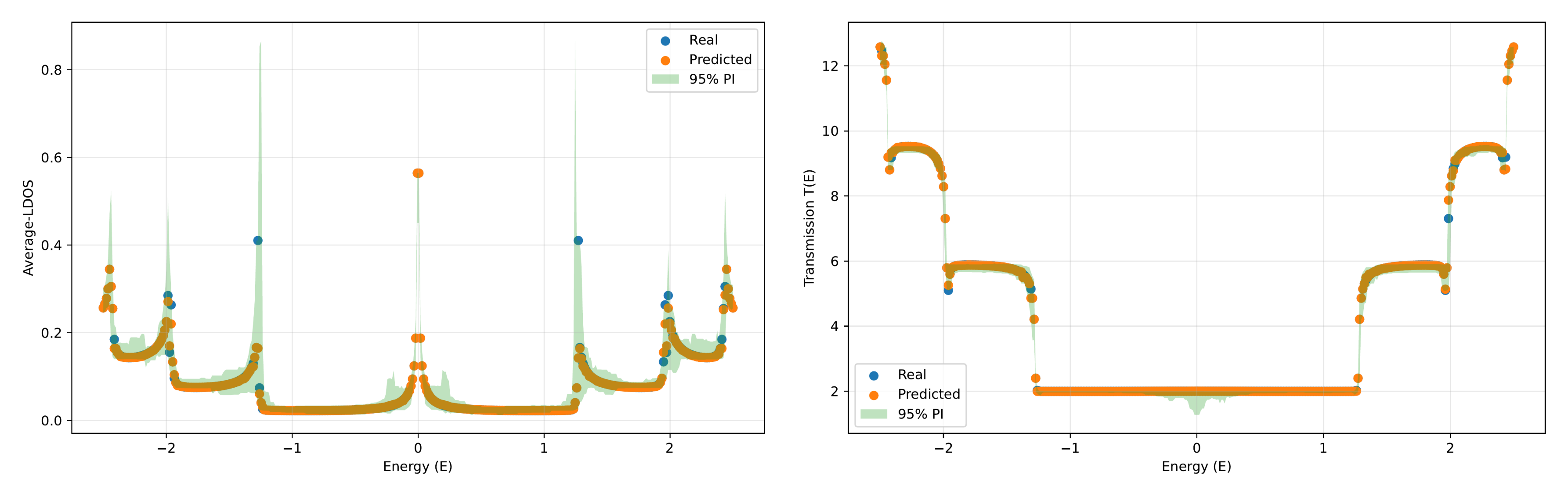}
    \caption{ Predicted (orange) and Real (blue) spectral curves for Average-LDOS and Transmission $T(E)$  as a function of energy ($E$), shown for one representative configuration from the test dataset. The green shaded region indicates the $95\%$ Prediction Interval (PI).}
    \label{fig:pre}
\end{figure}

The comparative performance of the regression model across the two target properties, Transmission  and Average-LDOS, is visualized in Fig \ref{fig:pi}, which plots the predicted values against the true values alongside the $ 95\% $  Prediction Interval (PI). The predictions for Transmission ($T$) exhibit exceptionally high precision, clustering tightly around the ideal $y=x$ line with a notably narrow $95\%$ PI across the full range of values . This visual agreement confirms the model's high predictive power for the global transport property.In contrast, the predictions for Average-LDOS show a marked increase in uncertainty and scatter, particularly evident in the wider and more volatile $95\%$ PI . Furthermore, a systematic bias emerges at higher Average-LDOS values (above $\approx 2.0$), where the model consistently underpredicts the true response. his increased prediction error for the local Average-LDOS property highlights its inherent complexity and greater sensitivity to fine geometric variations, emphasizing why approaches like polynomial feature expansion were necessary to achieve even this level of accuracy.
Fig.\ref{fig:pre}  complements this analysis by showing the full spectral prediction for both Average-LDOS and Transmission, $T(E)$, as a function of energy ($E$) for one random configuration in the test dataset. The fidelity of the predicted spectral curves (orange dots) relative to the real NEGF calculated values (blue dots) visually confirms the quantitative findings. The model successfully reproduces the intricate features of both $T(E)$ and Average-LDOS spectra, with the $95\%$ PI (green shaded area) accurately encompassing the real values at nearly all energy points.

To evaluate the models performance on extrapolation data, i.e., system configurations outside the training range, we generated a test dataset by graphene, stanene, silicene, and germanene; in the system shapes of: $w:14,36$ ; $L:5,9$ ;  $nm:6\%$ and predict targets with our models.  
\begin{table}
    \centering
     \resizebox{\textwidth}{!}{
    \begin{tabular}{|c|c|c|c|c|c|} \hline 
         \textbf{Target}&  \textbf{MAE}&  \textbf{RMSE}&  \textbf{$R^2$} &  \textbf{Accuracy}& \textbf{F1}\\ \hline 
         Average-LDOS &   0.028&  0.054&  0.931&  0.9717& 0.960\\ \hline 
         T&  0.847&  1.493&  0.917&  0.595& 0.586\\ \hline
    \end{tabular}
   }
    \caption{Performance of models to predict $T$ on extrapolation data}
    \label{tab:performance models}
    
\end{table}
\begin{figure}
    \centering
    \includegraphics[width=1\linewidth]{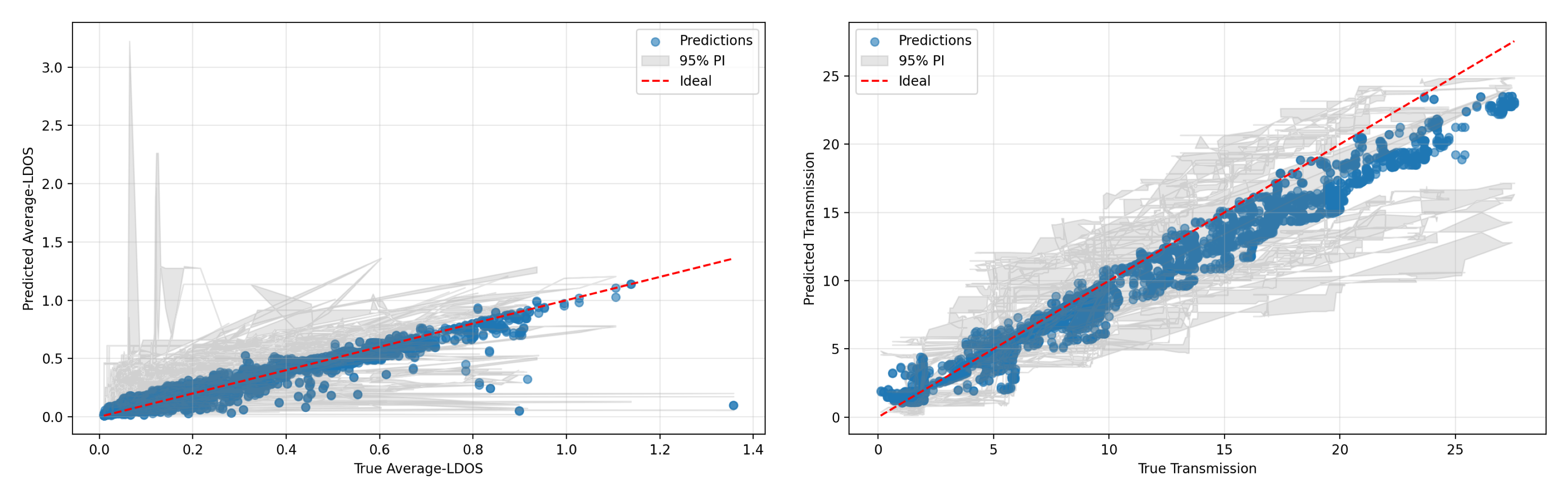}
    \caption{Regression model performance on extrapolation dataset. Comparing the predicted vs. true values for Average-LDOS and Transmission ($T$) . The dashed red line represents the ideal prediction ($y=x$), and the shaded area shows the $95\%$ Prediction Interval (PI).}
    \label{fig:expi}
\end{figure}
Table \ref{tab:performance models} presents the performance of our models on extrapolation data. Significant decreases in performance are observed: the accuracy of the RF regression model for 
$T$ drops by $38\%$, and for Average-LDOS, it drops by $25\%$. The evaluation of the model on the extrapolation dataset, visualized in Fig.\ref{fig:expi}, reveals the inherent limitations of generalization outside the training domain, particularly in the model's confidence. For Transmission, the model maintains reasonable accuracy along the $y=x$ line, demonstrating robustness for the global transport property even in highly disordered and short systems. However, the $95\%$ Prediction Interval (PI) significantly widens compared to the in-domain test, reflecting a substantial loss in prediction certainty as the model moves away from familiar configurations.Also for Average-LDOS, the extrapolation performance degrades.  The true Average-LDOS values in this test are in the low range ($\approx 0.0$ to $1.4$), a region which showed the highest certainty in the in-domain analysis. Nevertheless, PI becoming extremely volatile and widely dispersed across the entire plot and performance decrease due of novel geometric inputs.

\begin{figure}
    \centering
    \includegraphics[width=1\linewidth]{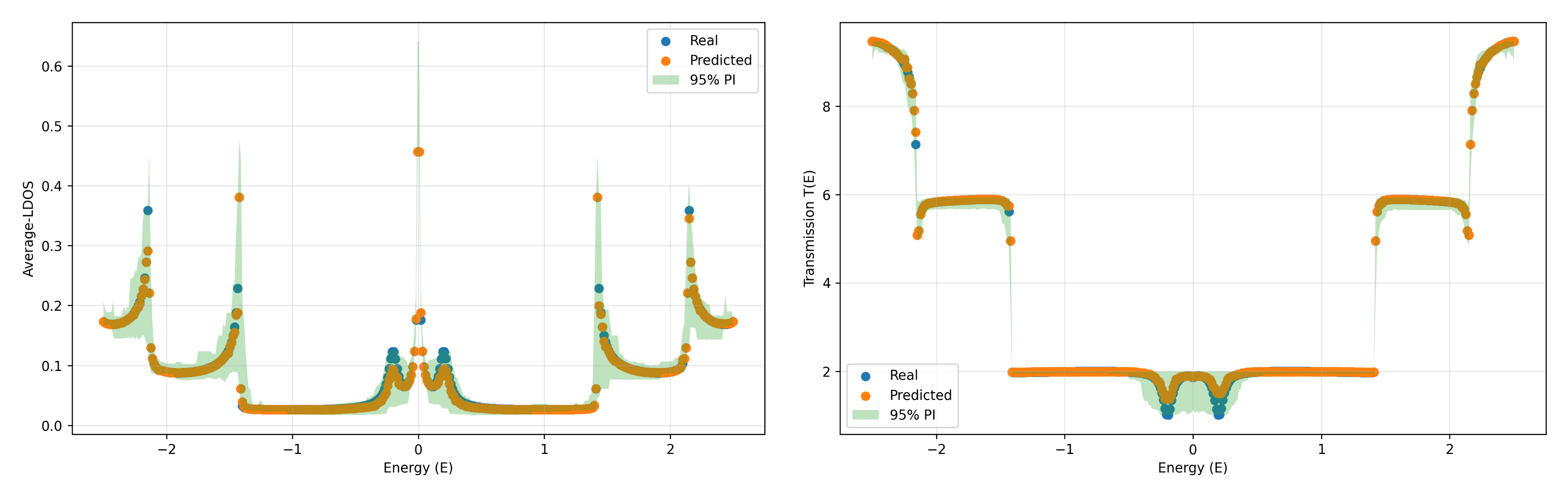}
    \caption{Predict $T$ and Average-LDOS values for extrapolation data of graphene system. }
    \label{fig:expre}
\end{figure}

The overall degradation in performance across all metrics highlights a fundamental limitation of RF models.
Fig.\ref{fig:expre} is the predicted values versus reals of graphene sample where shows that the Random Forest model struggles with extrapolation. While it performs well on training and test data, its accuracy drops significantly when applied to configurations outside the training range. This is because RF builds decision trees based on thresholds observed during training, and when features exceed these thresholds, the model lacks meaningful rules to guide predictions. As a result, it often defaults to predictions from boundary regions of the training set, leading to poor generalization.

Random Forest (RF) has been successfully applied in previous works to classify 
$T$, demonstrating its ability to capture transport patterns\cite{Ghosh2023-pc} \cite{9216557}.  However, such models typically focus on small-scale systems or limited feature sets, and do not address scalability across material types and device geometries. In this work, we extend the analysis by testing RF in both regression and classification settings while ensuring the model can generalize to larger system sizes and impurity levels. Importantly, we assess the models limitations by testing them on extrapolation data, a critical step often missing in similar works. These results clearly demonstrate both the practical value and boundaries of Random Forest models in quantum transport prediction.

\section{conclusion }

Here, we built data-driven and scalable machine learning models to predict key quantum transport quantities namely, the transmission coefficient $T(E)$ and the local density of states ($LDOS$) in hexagonal two-dimensional (2D) lattices with randomly located magnetic impurities. The model for the physical system was a tight-binding Hamiltonian, and the transport quantities were computed using the Non-Equilibrium Green's Function (NEGF) formalism. By constructing a dataset that spans a broad range of device geometries, impurity concentrations, and energy levels, we ensured that our models learned from physically valid and diverse configurations.
Features were selected based on the geometric and physical characteristics of the lattice;  including the width and length of the scattering region, number of impurities, hopping amplitude, and target energy. Our results reveal that Random Forest regression models are superior to classification-based approaches in accuracy, stability, and the ability to learn continuous variations in quantum transport behavior. Regression models possessed extremely low error measures and very high $R^2$
scores on test data,  successfully learning the underlying nonlinear relationships created by impurity-induced disorder and interference effects.

However, although these models were excellent on in-domain data, they degraded severely when evaluated on extrapolation scenarios, i.e., device configurations outside the training interval. This finding is particularly crucial for researchers wishing to deploy these models in actual applications where configurations unheard of elsewhere dominate.

The predictive accuracy of our method offers significant advantage in the high-throughput screening and optimization of 2D material devices, particularly in spintronics and nanoelectronics, where computationally costly calculations are frequent. By replacing expensive quantum transport calculations with cheap and accurate machine learning models, researchers can accelerate the design process and better use computational resources.

In the future, one can look forward to seeing further work in the direction of merging more expressive and adaptive learning frameworks such as Physics-informed neural networks (PINNs), graph neural networks, or Gaussian processes to improve generalization, especially in extrapolation regimes. Further, opening up the input feature space to include spin-orbit coupling, edge geometries, temperature effects, or even strain and external fields would help to further improve the physical accuracy and predictive power of the model. These developments would allow the models to be used for a broader class of materials and device geometries, ultimately pushing the limits of machine learning in quantum transport.

\section*{Data and Code Availability}
\noindent The machine learning code developed for this study and the dataset used to train and evaluate the models are publicly available in the \href{https://doi.org/10.5281/zenodo.17109064 }{Zenodo repository} \cite{seyed_mahdi_mastoor_2025_17109064}. The corresponding live repository for code development is available on \href{https://github.com/SMahdiMastoor/ML-Quantum-Transport-in-Disordered-2D-Hexagonal-Materials} {GitHub}

\section*{Conflicts of interest}
\noindent There are no conflicts to declare.
\section*{Author Contributions}
\noindent Seyed Mahdi Mastoor: investigation, validation, software, programming,
writing – original draft, writing – reviewing, editing, resources
and Amirhossein Ahmadkhan Kordbacheh: supervision, project administration, writing – reviewing, editing, investigation.

\bibliographystyle{elsarticle-num}
\bibliography{ref}
\end{document}